# Magneto-optical properties of $La_{2/3}Sr_{1/3}MnO_3$ ultrathin films

M. Veis[1][1], M. Zahradnik[1], R. Antos[1], S. Visnovsky[1], Ph. Lecoeur[2], D. Esteve[2], S. Autier-Laurent[2], J.-P. Renard[2], and P. Beauvillanin[2]

[1] Charles University of Prague, Faculty of Mathematics and Physics, Ke Karlovu 3, 12116 Prague 2, Czech Republic
[2] Institut d'Electronique Fondamentale, IEF/UMR 862, Université Paris-Sud XI, Bâtiment 220, 91405 Orsay Cedex, France

Abstract:
Pulse laser deposited $La_{2/3}Sr_{1/3}MnO_3$ ultrathin films on $SrTiO_3$ substrates were characterized by polar and longitudinal Kerr magneto-optical spectroscopy. An agreement between experimental and theoretical spectra was achieved for a 10.7 nm thick film, while a distinction in the Kerr effect amplitudes was obtained for a 5 nm thick film. This points to the slight suppression of ferromagnetism by interface effects. Nevertheless, the room temperature ferromagnetism of the 5 nm thick film was clearly demonstrated. All the studied films exhibit magneto-optical properties similar to single crystals and thick films, which confirms a fully developed perovskite structure even at 5 nm.

Hole doped manganites $La_{1-x}M_xMnO_3$ (M = Ca, Sr, Ba) with perovskite-type structure have been intensively studied in the last decades due to their unique physical properties. The colossal magnetoresistance effect[1] and extremely high degree of spin polarization make them promising candidates for applications in spintronics. Their structural, magnetic and electric properties are strongly correlated, and can be optimized by proper choice of substrate, doping level or deposition conditions. It was realized that the large number of degrees of freedom in manganites could be used to design their physical properties according to some specific function. This large variety and tunability of the physical properties (ferromagnetism, antiferromagnetism, metallicity, superconductivity, optical properties, etc.) provides fabulous advantages for various device applications. New devices, such as all-oxide spin FETs[2], magnetic tunnel junctions (MTJs)[3] and multiferroic memories[4], often include mixed-valence manganites as metallic contacts with well-defined magnetic ordering.

---

[1] Author to whom correspondence should be addressed. Electronic mail: veis@karlov.mff.cuni.cz.

Many investigations have been focused on La$_{2/3}$Sr$_{1/3}$MnO$_3$ (LSMO), which shows the highest Curie temperature among the family of manganites ($T_C \approx 370$ K)[5]. Furthermore, it exhibits almost 100% spin polarization[6] and the largest single electron bandwidth, which is very important from the application point of view. Ferromagnetic properties of LSMO are dominated by the double-exchange (DE) interaction, which originates from the $e_g$ electron transfer between Mn$^{3+}$ and Mn$^{4+}$ ions via the O$^{2-}$ *2p* state. Since the DE electron transfer probability strongly depends on Mn$^{3+}$-O-Mn$^{4+}$ geometry (changes in Mn-O bond lengths and/or Mn$^{3+}$-O-Mn$^{4+}$ bond angles), the main factors responsible for changes in magnetic properties are distortions of MnO$_6$ octahedra mainly induced by the strain arising in the film from the lattice mismatched substrate[7]. When grown on SrTiO$_3$ (STO) substrates (a = 0.3905 nm, cubic) a tensile strain is induced in the LSMO layer, resulting in an elongation of a$_{in-plane}$ lattice parameter with respect to the bulk value (a$_{LSMO\ bulk}$ = 0.3889 nm) and compression of c$_{out-of-plane}$ lattice parameter. The variation of the (c$_{out-of-plane}$/a$_{in-plane}$) ratio as a function of thickness was reported to be negligible[8], reflecting a fully strained state without any relaxation parameter between the interface and the surface up to 60 nm. It was shown that the three-dimensional compression and biaxial distortions affect $T_C$[7, 9]. Physical properties considerably different from bulk crystals were observed in LSMO ultrathin films[10, 11]. It was shown both theoretically and experimentally that a strain and LSMO/STO interface effects are responsible for a drastic change in the transport and magnetic properties[12, 13]. However, there are various explanations of the origin of these interface effects[12, 14-16]. Studying the physical properties of LSMO ultrathin films and optimizing their growth process is therefore necessary in order to avoid the negative influence of interface effects and to achieve high quality ultrathin films with well-defined physical properties suitable for spintronic applications.

Magneto-optical (MO) spectroscopy offers an opportunity to study the physical properties of magnetic materials when other conventional methods might not be effective. Previous MO studies were therefore focused on LSMO thin films[17-21], thick films[22], crystalline pellets[23] and single crystals[24]. However, since the MO Kerr effect is proportional to the magnetization $M$ in the film, ultrathin films exhibit extremely small angles of the Kerr rotation and ellipticity, requiring high sensitivity experimental MO setup. Therefore a systematic study of magneto-optical properties of LSMO films with thickness bellow 10 nm is still missing.

In this letter we report about magneto-optical properties of LSMO ultrathin films grown by pulsed laser deposition on single crystal (100) STO substrates. The films were deposited from a stoichiometric target under low oxygen pressure of 120 mTorr using a KrF laser at wavelength of 248 nm. The maximum energy fluence was 3 J/m$^2$ with the pulse repetition rate of 1 Hz and the substrate was heated to 620°C during the deposition. These parameters were found to be optimal for a single-crystalline film growth. Low pressure PLD under a strong oxidizing gas allows 'cell-by-cell' growth which can be monitored in situ using reflective high-energy electron diffraction and which produces high quality epitaxial films or superlattices[7]. The quality of interfaces along with smoothness of the surface was increased more by modification of the laser beam using a spatial filter, which acted as a beam homogenizer. The surface roughness was probed by atomic force microscopy and was found lower than 0.2 nm for all investigated films. The thickness, $t$, of LSMO layers was found 5 and 10.7 nm using X-ray diffraction (XRD) as well as X-ray reflectivity measurements. Out of plane parameters, $c$, of LSMO layers were obtained from XRD spectra using (002) LSMO peak. The values are $c = 0.3845$ nm for 10.7 nm thick sample and $c = 0.3833$ nm for 5 nm thick sample, indicating fully strained LSMO layers. The superconducting quantum interference device (SQUID) magnetometry showed $T_C$ above the room temperature for both samples, reflecting good quality of LSMO layers.

The magneto-optical spectroscopy was carried out using an azimuth modulation technique with synchronic detection in polar and longitudinal configuration. The experiment has been done in the photon energy range between 1.2 and 4.6 eV The experimental optical set up included a 450W high power Xe arc lamp, quartz prism monochromator, polarizer, dc compensating Faraday rotator, ac modulating Faraday rotator, phase plate (for Kerr ellipticity measurements), sample in magnetic field, analyzer and photomultiplier. In the small angle approximation the complex polar MO Kerr effect was measured at nearly normal light incidence as a ratio $\theta_K + i\varepsilon_K \approx (r_{yx}/r_{xx})$ of Jones reflection matrix elements, where $\theta_K$ and $\epsilon_K$ are the Kerr rotation and ellipticity. The longitudinal MO Kerr effect was measured similarly at the angle of incidence adjusted to 56° for p-polarized incident light as a ratio $\theta_K + i\varepsilon_K \approx (r_{ps}/r_{pp})$ [17]. Applied magnetic field was 470 mT and 100 mT in the polar and longitudinal configuration, respectively. In the both configurations the magnetic field was sufficient for the film saturation (as was checked by the measurement of magnetic field dependence of the MO Kerr effect). During the polar configuration measurements the samples were placed on a water-cooled pole piece of electromagnet and their temperature was stabilized at 285 K. In the longitudinal configuration measurements the samples were kept at the stabilized room temperature of 295 K. Theoretical calculations were performed using the transfer matrix approach and assuming the film uniformity. Since the back-unpolished side of the STO substrate was depolarizing and therefore had a negligible contribution to the optical reflection, we focused ourselves to a model of a single layer on semi-infinite substrate. For strained LSMO layers this model (assuming flat surface and planar interface between the LSMO layer and STO substrate) represents only an approximation.

Figure 1 displays experimental polar Kerr rotation and ellipticity spectra for the sample with the thickness $t = 10.7$ nm. Two opposite spectroscopic peaks (centered near 2.7 and 3.6 eV) dominate the Kerr rotation spectrum and one positive peak (centered near 3.4 eV) dominates the Kerr ellipticity spectrum. The spectra have similar spectral behavior with previously published magneto-optical studies on thin films[15, 17, 18], thick films[19], single crystals[24] and bulk crystalline pellets[23]. A splitting of the negative peak in the Kerr rotation spectrum is clearly visible near 3.7 eV. This effect was observed only for thin layers[15, 17, 18]. It originates from the STO substrate, as was found by spectroscopic ellipsometry (SE) measurements on bare STO crystals, and can be seen thanks to a reduced penetration depth, which is approximately 30 nm at this photon energy[17]. The smoothness of the experimental spectra (without any noise) demonstrates high optical quality of the surface. The obtained spectra exhibit higher amplitudes than previously published results on LSMO films with similar thickness[15, 17, 18]. This is pointing to better magnetic properties of the investigated film, which were achieved by a homogenized laser beam.

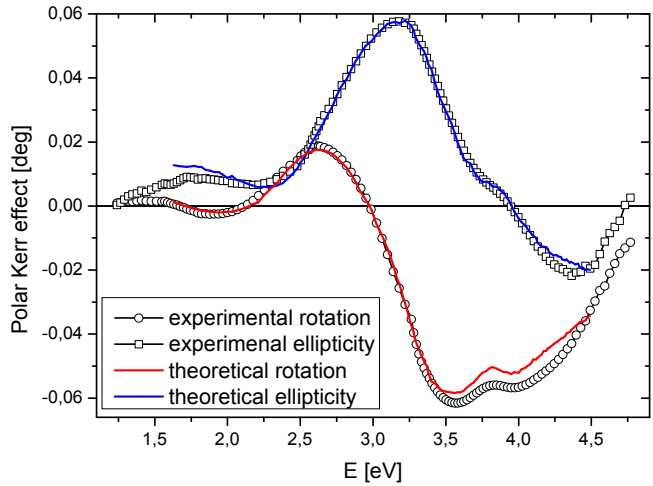

Figure 1. (Color online) Experimental (symbols) and theoretical (solid curves) polar Kerr rotation and ellipticity spectra of 10.7 nm thick LSMO film. The experimental spectra were obtained for nearly normal light incidence. The theoretical model employed transfer matrix formalism and material parameters of a 35 nm thick LSMO layer.

Theoretical models are included in Figure 1. Optical and magneto-optical constants (diagonal and off-diagonal elements of permittivity tensor $\varepsilon_{xx}$ and $\varepsilon_{xy}$) obtained on a 35 nm thick LSMO layer on STO (100) substrate[17] were adopted in this calculation. Optical constants of STO substrate (diagonal elements of permittivity tensor $\varepsilon_{xx}$) used in the calculation were acquired from SE measurements on bare STO substrates[18]. An excellent agreement between experimental data and theoretical calculations is clearly visible. A small deviation in the UV region can be explained by slightly better functionality of the current magneto-optical setup, which now employs second photomultiplier with enhanced UV sensitivity. The broadening of the Kerr rotation at 3.7 eV is also well reproduced since the layer thickness and the back reflection from the layer/substrate interface is included in the calculation procedure. The agreement between the theoretical and experimental spectra indicates well defined and flat STO/LSMO interface since the model structure consider planar interfaces without any change of magnetic properties throughout the layer.

It appears that interfacial effects between the LSMO layer and STO substrate, which are the reason for magnetization lost at the interface, has an insignificant influence to the magneto-optical properties in the investigated film (i.e. they occur at the short distance in the proximity of the interface). A suppression of these effects is desirable for device applications with regard to their negative influence on the properties of MTJs. Their origin has been recently extensively discussed in literature[12-14, 25-27]. Magnetic resonance measurements and scanning tunneling microscopy[28] have shown that these effects are connected to a phase-separation phenomenon[29] at the interface where ferromagnetic insulating and metallic phases separate at a scale of a few nanometers[30]. The origin of such phase separation can be related to the presence of structural inhomogeneities localized at the interface between the film and substrate[31]. Tebano *et al.*[13] reported that the origin of the interface effects is related to the suppression of DE mechanism due to the broken symmetry at the interfaces, which stabilizes the $(3z^2-r^2)$ orbitals against $(x^2-y^2)$[13] making a local C-type antiferromagnetic structure at the interface. This effect is driven by the strain, which is induced by the STO substrate. On the other hand a fingerprint of the $(x^2-y^2)$ preferential orbital ordering was observed by linear dichroism of the x-ray absorption (LD-XAS) for films with the thickness 20 nm (50 unit cells). Such orbital character was even more visible in relatively thick (100 unit cells, roughly 40 nm) fully strained LSMO/STO films (c/a = 0.98). These films displayed metallic ferromagnetic properties with $T_C \approx 370$ K identical to the bulk value[13]. Magneto-optical properties of such thick films (t ≥ 35 nm) are comparable with those of bulk crystals[17, 18]. On the other hand Huijben *et al.*[16] observed by LD-XAS that preferred orbital ordering remains $(x^2-y^2)$ for film thickness down to few unit cells.

Since the material parameters of the 35 nm thick layer provided an excellent agreement between the experiment and theoretical calculation we can conclude that the magneto-optical properties of 10.7 nm thick film are close to those of bulk crystals. With respect to the report of Tebano *et al.*[13, 14] our results suggest that the preferred ($x^2$-$y^2$) orbital ordering should be responsible for magneto-optical properties even if the thickness of LSMO layer is only 10.7 nm (27 unit cells). This is in agreement with observations of Huijben *et al.*[16] and theoretical explanations given by Lepetit *et al.*[12] who proposed preferred ($3z^2$-$r^2$) ordering only in the proximity of the interface (in the thickness of 3 unit cells).

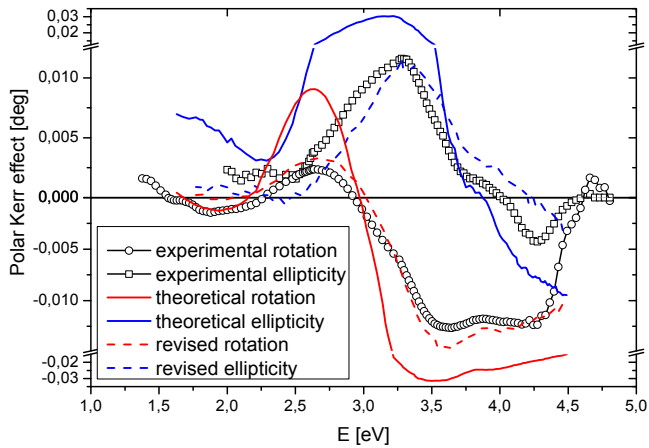

Figure 2. (Color online) Experimental (symbols) and theoretical (solid and dashed curves) polar Kerr rotation and ellipticity spectra of 5 nm thick LSMO film. The experimental spectra were obtained for nearly normal light incidence. The theoretical model (solid lines) employed transfer matrix formalism and material parameters of a 35 nm thick LSMO layer. The revised theoretical model used material parameters of a 22 nm thick LSMO layer.

Figure 2 shows experimental polar Kerr rotation and ellipticity spectra for 5 nm thick film. Similarly to the thicker film the spectral behavior contains spectroscopic features typical for LSMO single-crystalline layers. This gives a clear evidence of fully developed LSMO structure even at such low thickness, which shows ferromagnetic behavior at room temperature. This is in contrast with previously published results[12, 13], where $T_C$ was reported lower than 280 K. Room temperature ferromagnetism in 5 nm thick LSMO layer suggests that the LSMO/STO interfacial layer has indeed the thickness of only a few unit cells[12, 27]. Moreover, the low level of noise in the spectra reflects a good optical quality of the surface. Theoretical spectra were calculated using the same material parameters as for the previous sample and are included (as solid curves) in Fig. 2. Owing to results from SQUID measurements and previous studies[12, 13, 16], which reported suppressed ferromagnetism in LSMO layers of the thickness bellow 18 unit cells with respect to bulk crystals, we should observe a difference between theoretical and experimental polar Kerr spectra. Indeed, although the theoretical spectral dependence follows the experimental data reasonably, there is an evident difference in their amplitudes. Therefore a new model which employed material parameters deduced from 22 nm thick LSMO layer[17] has been calculated and the spectra are also included (as dashed curves) in Fig. 2. A reasonable agreement between experimental data and revised theoretical spectra is clearly visible. The spectral dependence of the experimental data is well reproduced by theoretical calculation and deviations can be addressed to the slight difference in optical properties between 5 and 22.3 nm thick LSMO layer since their deposition conditions were different. Therefore the preferred orbital ordering in the layer still remains ($x^2$-$y^2$), but a fingerprint of ($3z^2$-$r^2$) occupation is noticeable as the decrease of Kerr effect amplitudes. This is understandable if one considers the same thickness of the interfacial layer and the reduced overall thickness of LSMO layer. While the volume of the ($3z^2$-$r^2$) preferred orbital ordering in the interfacial LSMO/STO layer remains same, the volume of ($x^2$-$y^2$) preferred orbital ordering is decreased by smaller film thickness. This leads to the decrease of DE interactions, which are responsible for ferromagnetic properties in LSMO and consequently to the decrease of MO Kerr effect amplitudes.

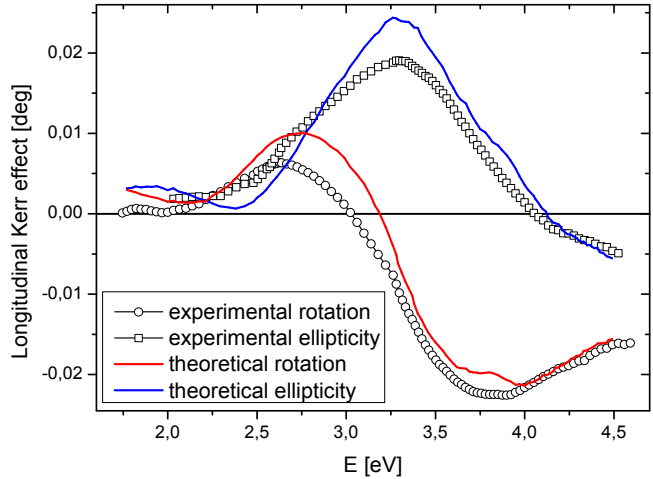

Figure 3. (Color online) Experimental (symbols) and theoretical (solid curves) longitudinal Kerr rotation and ellipticity spectra of 10.7 nm thick LSMO film. The angle of light incidence was adjusted to 56°. The incident light was *p*-polarized. The theoretical spectra were calculated similarly to the polar case.

Experimental Kerr rotation and ellipticity spectra measured in longitudinal configuration together with theoretical models for 10.7 nm thick LSMO layer are shown in Fig. 3. The spectra are similar in shape with previously published results[17]. The theoretical model describes the experimental spectra reasonably and the deviations can be assigned to a higher temperature during longitudinal measurements (approx. 10 K higher), because the ferromagnetic properties of LSMO are strongly temperature dependent near $T_C$. Since the magneto-optical spectra depend on the magnetization *M*, the higher temperature of the sample results in smaller amplitudes of the magneto-optical effect.

Experimental and theoretical longitudinal Kerr rotation spectra of 5 nm thick LSMO layer are displayed in Fig. 4. Despite a higher level of noise a spectral dependence of the longitudinal Kerr rotation is noticeable and similar to that of the thicker sample. The higher level of noise is due to a very small Kerr effect amplitude, which is in order of millidegrees. Such value is at the edge of the experimental system sensitivity. The theoretical spectral dependence is in agreement with the experimental spectrum. Notable differences in amplitudes in UV region can be assigned to limited sensitivity of magneto-optical setup for such small angles. However, the visible longitudinal spectrum of 5 nm thick LSMO layer is a good demonstration of magneto-optical setup efficiency for probing magnetic properties of nanostructures.

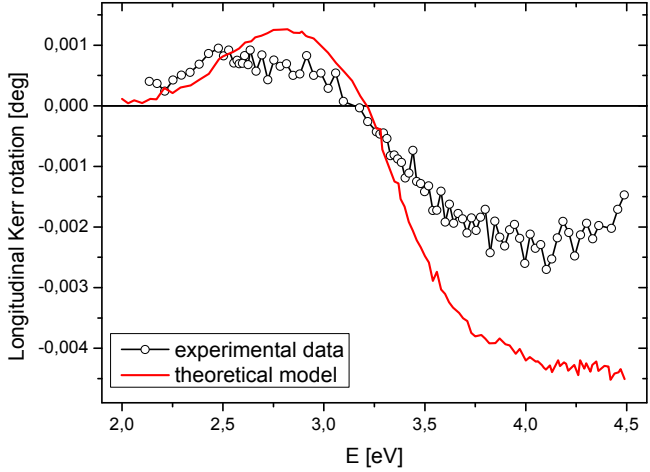

Figure 4. (Color online) Experimental (symbols) and theoretical (solid curve) longitudinal Kerr rotation spectra of 5 nm thick LSMO film. The angle of light incidence was adjusted to 56°. The incident light was *p*-polarized. The theoretical spectrum was calculated similarly to the polar case. Material parameters of a 22 nm thick LSMO layer were used in the calculation.

In summary, we have reported about magneto-optical properties of ultrathin LSMO films grown on STO substrates. All films displayed spectral properties of the magneto-optical Kerr effect similar to those of bulk single crystals. Bulk-like magneto-optical properties of 10.7 nm thick film were confirmed by excellent agreement between experimental data and theoretical calculations. This suggests preferred ($x^2$-$y^2$) ordering in the layer. Smooth polar Kerr spectra of 5 nm thick film confirmed fully developed LSMO structure. Suppression of Kerr effect amplitudes pointing on the possible fingerprint of ($3z^2$-$r^2$) orbital ordering, which is caused by the interface effects between the layer and the substrate, and affects DE interactions responsible for ferromagnetism in LSMO. Bulk-like magneto-optical properties of the 10.7 nm thick film and ferromagnetic behavior of the 5 nm thick film at the room temperature demonstrated a perfect control of the layer by layer growth mode which leads to fully strained films. This could be obtained by a perfect control of the laser beam with top hat profile achieved by the beam homogenizer. Such films are suitable for spintronic applications. Furthermore, visible spectrum of longitudinal Kerr rotation of 5 nm thick film clearly demonstrated that the magneto-optical spectroscopy is highly effective experimental method to probe physical properties of magnetic nanostructures.

Acknowledgements

This work was supported by Czech Science Foundation grant no. P204/10/P346.